\documentclass[onecolumn,showpacs,11pt]{revtex4}
\usepackage{graphicx}
\usepackage{dcolumn}
\usepackage{bm}
\usepackage{epstopdf}
\begin{document}
%%%%%%%%%%%%%%%%%%%%%%%%
\newcommand{\hs}{\hspace*{0.3cm}}
\newcommand{\vs}{\vspace*{0.3cm}}
\newcommand{\be}{\begin{equation}}
\newcommand{\ee}{\end{equation}}
\newcommand{\bea}{\begin{eqnarray}}
\newcommand{\eea}{\end{eqnarray}}
\newcommand{\ben}{\begin{enumerate}}
\newcommand{\een}{\end{enumerate}}
\newcommand{\bde}{\begin{widetext}}
\newcommand{\ede}{\end{widetext}}
\newcommand{\nn}{\nonumber}
\newcommand{\crn}{\nonumber \\}
\newcommand{\Tr}{\mathrm{Tr}}
\newcommand{\non}{\nonumber}
\newcommand{\noi}{\noindent}
\newcommand{\al}{\alpha}
\newcommand{\la}{\lambda}
\newcommand{\bet}{\beta}
\newcommand{\ga}{\gamma}
\newcommand{\va}{\varphi}
\newcommand{\om}{\omega}
\newcommand{\pa}{\partial}
\newcommand{\+}{\dagger}
\newcommand{\fr}{\frac}
\newcommand{\bc}{\begin{center}}
\newcommand{\ec}{\end{center}}
\newcommand{\Ga}{\Gamma}
\newcommand{\de}{\delta}
\newcommand{\De}{\Delta}
\newcommand{\ep}{\epsilon}
\newcommand{\varep}{\varepsilon}
\newcommand{\ka}{\kappa}
\newcommand{\La}{\Lambda}
\newcommand{\si}{\sigma}
\newcommand{\Si}{\Sigma}
\newcommand{\ta}{\tau}
\newcommand{\up}{\upsilon}
\newcommand{\Up}{\Upsilon}
\newcommand{\ze}{\zeta}
\newcommand{\ps}{\psi}
\newcommand{\Ps}{\Psi}
\newcommand{\ph}{\phi}
\newcommand{\vph}{\varphi}
\newcommand{\Ph}{\Phi}
\newcommand{\Om}{\Omega}
%%%%%%%%%%%%%%%%%%%%%%%%
\title{Inverted neutrino mass hierarchy and mixing in the Zee-Babu model}

\author{Vo Van Vien}
\email{wvienk16@gmail.com} \affiliation{Department of Physics, Tay
Nguyen University, 567 Le Duan, Buon Ma Thuot, DakLak, Vietnam}

\author{Hoang  Ngoc  Long}
\email{hnlong@iop.vast.ac.vn} \affiliation{Institute of Physics,
VAST, 10 Dao Tan, Ba Dinh, Hanoi, Vietnam}

\author{Pham Ngoc Thu}
\email{phamngocthutb@gmail.com} \affiliation{Faculty of
Mathematics-Physics-Informatics, Tay Bac University, Son La city,
Vietnam}

\date{\today}

\begin{abstract}

We show that the neutrino mass matrix of the Zee-Babu model is
able to fit the recent data on neutrino masses and mixing
with non-zero $\theta_{13}$ in the inverted neutrino mass hierarchy. The results show that the Majorana  phases are equal to zero and the Dirac phase ($\de$) is
predicted to either $0$ or $\pi$, i. e, there is no CP violation in the Zee-Babu model at the two loop level. The effective mass governing
neutrinoless double beta decay and the sum of neutrino masses are
consistent with the recent analysis.
\end{abstract}


\maketitle

%\tableofcontents

\section{\label{intro}Introduction}

At present, neutrino and Higgs physics are hot topics in current
Particle Physics. The neutrino mass and mixing are the first
evidence of beyond Standard Model (SM) physics. Despite the Higgs
boson bas been discovered by the ATLAS \cite{ATLAS} and the CMS \cite{CMS} but in which
model it belongs is still open question. For the aforementioned
reasons, the search for an extended model coinciding with the
current data on neutrino physics is one of  our top priorities. In
our opinion, the model with the simplest particle content is
preferred. By this criterion,  the Zee-Babu model
\cite{zee,zeebabu,babu}
 is very attractive. In our previous work \cite{lvZB}, we have
 derived the exact solution for the neutrino mass matrix in the
 model under consideration and derived some regions of the
 parameters in the normal neutrino mass hierarchy.

As far as we know at present the values of the
absolute neutrino masses as well as the mass ordering of
neutrinos are still an open problem. The mass
ordering of neutrino depends on the sign of $\Delta m^2_{31}$
which is currently unknown. In the case of 3-neutrino mixing, the two possible signs of $\Delta
m^2_{31}$ corresponding to two types of neutrino mass spectrum can
be provided as follows

\ben \item Normal hierarchy (NH): $
|m_1|\simeq |m_2| < |m_3|,\,\, \Delta m^2_{31}=m^2_3-m^2_1>0.$
\item Inverted hierarchy (IH): $|m_3|< |m_1|\simeq |m_2|,\,\, \Delta m^2_{31} =m^2_3-m^2_1<0$.\een

In this paper, we focus on the
 effective mass governing neutrinoless double beta decay and
the sum of neutrino masses. As will be discussed below, the model can give some regions of the parameters where neutrino mixing angles and
the inverted neutrino mass hierarchy obtained consistent with the
recent experimental data. Indeed, by starting from the neutrino mass matrix in the Zee-babu model\cite{zee,zeebabu,babu}, we get the
exact solution, i.e., the eigenstates and the eigenvalues. Comparing the model results with the experimental data we get the model parameters. The effective mass
governing neutrinoless double beta decay and the sum of neutrino
mass  are consistent with the recent analysis.

This paper is organized as follows. In Sec. \ref{model}, we briefly present the
Zee-Babu  model
and its neutrino mass matrix. Sec. \ref{solut} is devoted for the solution and
phenomenology with focus on the inverted spectrum.
We summarize our result  in the last section - Sec. \ref{conc}.

\section{Zee-Babu model and  neutrino mass matrix}
 \label{model}

With just two $\mbox{SU}(2)_L$ singlet Higgs fields, a singly
charged field $h^-$ and a doubly charged field $k^{--}$ and
without right-handed neutrinos, the new Yukawa interactions in the
Zee-Babu model \cite{zeebabu} are \be {\cal L}_Y =
f_{ab}\overline{(\psi_{aL})^C}\psi_{bL}h^+ +
h^\prime_{ab}\overline{(l_{aR})^C}l_{bR}k^{++}+H.c.,\label{pt151}
\ee where $\psi_L$ stands for the left-handed lepton doublet,
$l_R$ for the right-handed charged lepton singlet and ($a, b = e,
\mu, \tau$) being the generation indices, a superscript $^C$
indicating charge conjugation. Note that $f_{ab}$ is antisymmetric
($f_{ab}=-f_{ba}$) and $h^\prime_{ab}$ is symmetric
($h^\prime_{ab}=h^\prime_{ba}$).  In terms of the component
fields, the interaction Lagrangian is given by\bea
 {\cal L}_Y&=&2\left[ f_{e\mu}(\bar{\nu_e^c}\mu_L - \bar{\nu_\mu^c}e_L) +
  f_{e\tau}(\bar{\nu_e^c}\tau_L - \bar{\nu_\tau^c}e_L) +
   f_{\mu \tau}(\bar{\nu_\mu^c}\tau_L - \bar{\nu_\tau^c}\mu_L)\right]h^+\crn
&+&\left[h_{ee}\bar{e^c}e_R + h_{\mu \mu}\bar{\mu^c}\mu_R +h_{\tau
\tau} \bar{\tau^c}\tau_R + h_{e \mu}\bar{e^c}\mu_R + h_{e
\tau}\bar{e^c}\tau_R
+ h_{\mu \tau}\bar{\mu^c}\tau_R\right]k^{++} \label{pt151t}\\
&+&H.c. \nn\eea where we have used $h_{aa} = h^\prime_{aa}, h_{ab}
= 2 h^\prime_{ab} $ for $a \neq b$. In  Eq. (\ref{pt151}), the
lepton number is   conserved, and  neutrino mass will be generated
due to the Higgs potential given by: \be
V(\phi,h^+,k^{++})=\mu(h^- h^- k^{++}+h^+ h^+
k^{--})+\cdot\cdot\cdot.\label{pt152}\ee Here, the lepton number
is violated by two units, hence one expects the  Majorana neutrino
masses. From  Eq.(\ref{pt151}), it follows both $h^-$ and $k^{--}$
carry lepton number two, so the coefficient $\mu$ in (\ref{pt152})
also carries lepton number two. Therefore it is expected that the
Majorana neutrino masses are generated by loop quantum effects. At
the two-loop level, the  mass matrix  for Majorana neutrinos is
given by \be M_{ab} = 8\mu
f_{ac}h^*_{cd}m_cm_dI_{cd}(f^+)_{db},\label{pt152t}\ee where
$I_{cd}$ has the form \cite{tinhI}
 \bea
I_{cd}&=&\int\fr{d^4k}{(2\pi)^4}\int\fr{d^4q}{(2\pi)^4}\fr{1}{k^2-m^2_c}
\fr{1}{k^2-M^2_h}\fr{1}{q^2-m^2_d}\crn
&\times&\fr{1}{q^2-M^2_h}\fr{1}{(k-q)^2-M^2_k}. \label{pt153}\eea
Assuming masses of new Higgses are much larger than lepton ones,
we can  evaluate $I_{cd}$ as follows \be I_{cd} \simeq I =
\frac{1}{(16\pi^2)^2}\frac{1}{M^2} \fr{\pi^2}{3} \tilde{I}(r), \,
\, M \equiv  \textrm{max}(M_k,M_h). \label{pt154}\ee Here
$\tilde{I}(r)$ is a function of the ratio of the masses of the
charged Higgses $r \equiv M^2_k/M^2_h$,
\be \tilde{I}(r) =\left\{%
\begin{array}{ll}
     \hbox{$1 + \fr{3}{\pi^2}(\log^2 r - 1)$} & \mbox{for} \, \, r \gg 1 \\
\hbox{$1$} & \mbox{for} \, \,  r \rightarrow 0, \\
\end{array}%
\right.\label{lv1}\ee which is close to $1$ for a wide range of
scalar masses.

The neutrino mass matrix in (\ref{pt152t}) is symmetric
and given by \cite{babum}\bea &&{\cal M}_{\nu} = - I \mu f_{\mu \tau}^2
\times\crn
 &&\left(%
\begin{array}{ccc}
  \ep^2 \om_{\tau \tau } + 2 \ep \ep^\prime \om_{\mu \tau} + \ep^{\prime 2} \om_{\mu \mu}\,\,\,
   & \ep \om_{\tau \tau} +  \ep^\prime( \om_{\mu \tau} - \ep  \om_{e \tau}
 -\ep^{\prime } \om_{e \mu}) &\,\,\,-\ep^\prime \om_{\mu \mu}  -  \ep( \om_{\mu \tau}+  \ep \om_{e\tau}
 +   \ep^\prime \om_{e\mu}) \\
  \star &  \om_{\tau \tau} + \ep^{\prime 2} \om_{e e} -2\ep^\prime \om_{e\tau} &
   \ep \ep^\prime \om_{ee} - \om_{\mu\tau}-\ep \om_{e\tau}  +\ep^\prime \om_{e\mu} \\
  \star & \star &  \om_{\mu \mu} +2 \ep \om_{e\mu} + \ep^2 \om_{ee} \\
\end{array}%
\right)\crn \label{ktra}\eea where we have redefined parameters:
\be \ep \equiv \fr{f_{e\tau}}{f_{\mu \tau}}, \hs  \ep^\prime
\equiv \fr{f_{e\mu}}{f_{\mu \tau}} \hs \om_{a b} \equiv m_a
h^*_{ab} m_b .\label{ef}\ee
Let us denote \cite{ageng,lvZB} \bea \om_{\tau\tau}^{'} &\equiv&
 \om_{\tau\tau}
+ \ep^{\prime 2} \om_{e e} -2\ep^\prime \om_{e\tau},
 \crn
\om_{\mu\tau}^{'} &\equiv&   \om_{\mu\tau}+\ep
\om_{e\tau}  -\ep^\prime \om_{e\mu} -\ep \ep^\prime \om_{ee} , \crn
\om_{\mu\mu}^{'} &\equiv& \om_{\mu \mu} +2 \ep \om_{e\mu} + \ep^2
\om_{ee}\nn, \eea then the neutrino mass matrix can be rewritten
in the compact form
 \bea
 {\cal M}_{\nu} = - I \mu f_{\mu \tau}^2\left(%
\begin{array}{ccc}
  \ep^2 \om_{\tau \tau }^\prime + 2 \ep \ep^\prime \om_{\mu \tau}^\prime + \ep^{\prime 2} \om_{\mu \mu}^\prime
   & \ep  \om^\prime_{\tau \tau} +  \ep^\prime \om^\prime_{\mu \tau}  & -\ep  \om_{\mu\tau}^{\prime}
   -\ep^\prime \om_{\mu\mu}^{\prime} \\
  \star & \om_{\tau\tau}^{\prime} &
-   \om_{\mu\tau}^{\prime}  \\
  \star & \star &  \om^\prime_{\mu \mu} \\
\end{array}%
\right).\label{ktra}\eea
Next we turn to  solution and  implication to current neutrino data with a rather large $\theta_{13}$.

\section{Solution and phenomenology}
\label{solut} To begin this section, let us present the recent data on neutrino mass and mixing. The best fit values of neutrino
mass squared differences and the leptonic mixing angles in
\cite{Nuexp1} have been given to be slightly deviation
from Tri-bimaximal mixing form in the inverted spectrum, as shown in Tab. \ref{InvertedH} with a rather large $\theta_{13}$.
%%%%%%
\begin{table}[h]
\caption{\label{InvertedH}The experimental values of neutrino mass
squared splittings and leptonic mixing parameters, taken from
\cite{Nuexp1} for inverted hierarchy.} \bc
\begin{tabular}{lllllll}
\hline\noalign{\smallskip}
Parameter & Best fit & $1\sigma $ range& $2\sigma $ range \\
\noalign{\smallskip}\hline\noalign{\smallskip}
$\De m_{21}^{2}$($10^{-5}$eV$^2$) & $7.62$ & $7.43-7.81$ & $7.27-8.01$ \\
$\De m_{13}^{2}$($10^{-3}$eV$^2$)& $2.43$ & $2.37-2.50$ & $2.29-2.58$ \\
$\sin ^{2}\theta _{12}$ &  $0.32$ & $0.303-0.336$ & $0.29-0.35$  \\
 $\sin ^{2}\theta _{23}$&  $0.60$ & $0.569-0.626$ & $0.39-0.65$ \\
 $\sin ^{2}\theta_{13}$& $0.025$ & $0.0223-0.0276$ & $0.02-0.03$\\
\noalign{\smallskip}\hline
\end{tabular}
\ec
\end{table}

The matrix ${\cal M}_{\nu}$ in
(\ref{ktra}) has three exact eigenvalues given by \bea
\la_1&=&0,\crn
 \la_{2,3}&=&\frac{1}{2}\left(-k F \pm\sqrt{k^2\left[F^2+4(1+\epsilon^2+
 \epsilon'^2)(\om'^2_{\mu\tau}-\om'_{\mu\mu}
 \om'_{\tau\tau})\right]}\right),\label{l123}\eea
 where we have denoted
 \be
k=\mu If^2_{\mu\tau},\hs F=(1 +
\epsilon'^2)\om'_{\mu\mu}+2\epsilon\epsilon'\om'_{\mu\tau} + (1
+\epsilon^2)\om'_{\tau\tau}.\label{kF} \ee The massless
eigenstate  is given by \bea \nu_1 &=&
\fr{1}{\sqrt{f^2_{e\mu}+f^2_{e\tau}+f^2_{\mu\tau}}}
(f_{\mu\tau}\nu_e-f_{e\tau}\nu_\mu+f_{e\mu}\nu_\tau).\nn\\
\label{eigz}\eea Until now values of neutrino masses (or the
absolute neutrino masses) as well as the mass ordering of
neutrinos are unknown. An upper bound on the absolute value of
neutrino  mass was found from the analysis of the cosmological
data \cite{Tegmark} \be m_i\leq 0.6\, \mathrm{eV},\label{upb} \ee
 while the upper limit on the sum of neutrino mass is given in Ref. \cite{Ade}
\be \sum^{3}_{i=1} m_i\leq 0.66\,  \mathrm{eV}.
\label{upbsum}\ee
In the inverted hierarchy, three neutrino masses are chosen as follows: \be m_1=\la_3, \,\,
m_2=\la_2, \,\, m_3=0,\label{numassI}\ee with $\la_i \,\,\,
(i=1,2,3)$ is defined in (\ref{l123}), and the corresponding
eigenstates put in the neutrino mixing matrix: \bea U_{\nu
I}&=&\left(\begin{array}{ccc}
  \frac{A_2}{\sqrt{1+A^2_2+B^2_2}} & -\frac{A_1}{\sqrt{1+A^2_1+B^2_1}} &
  \frac{1}{\sqrt{1+\epsilon^2+\epsilon'^2}} \\
 \frac{B_2}{\sqrt{1+A^2_2+B^2_2}} &  - \frac{B_1}{\sqrt{1+A^2_1+B^2_1}} &
 -\frac{\epsilon}{\sqrt{1+\epsilon^2+\epsilon'^2}} \\
 \frac{1}{\sqrt{1+A^2_2+B^2_2}} &- \frac{1}{\sqrt{1+A^2_1+B^2_1}} &
 \frac{\epsilon'}{\sqrt{1+\epsilon^2+\epsilon'^2}}
\end{array}\right),\label{UnuI}
\eea where \cite{lvZB}\bea
A_{1,2}&=&\frac{-k\left[\epsilon(\epsilon'^2-1)\om'_{\mu\mu}+2\epsilon'(1+\epsilon^2)
\om'_{\mu\tau}+\epsilon(1+\epsilon^2)\om'_{\tau\tau}\right]
\pm\ep\sqrt{k^2F'}}
{2k\left[\epsilon\epsilon'\om'_{\mu\mu}+(1+\epsilon^2)\om'_{\mu\tau}\right]},\label{gta}\\
B_{1,2}&\equiv
&\frac{k(1+\epsilon'^2)\om'_{\mu\mu}-k(1+\epsilon^2)\om'_{\tau\tau}\pm\sqrt{k^2F'}}
{2k\left[\epsilon\epsilon'\om'_{\mu\mu}+(1+\epsilon^2)\om'_{\mu\tau}\right]},
\label{gtb} \eea and
% where
 \be
F'=F^2+4(1+\epsilon^2+\epsilon'^2)(\om'^2_{\mu\tau}-\om'_{\mu\mu}
 \om'_{\tau\tau}).\label{Fp}
\ee
The eigenstates  $\nu_i$ corresponding to the  eigenvalues
$m_i \, (i=1,2,3)$ are found to be \bea \nu_1&=&
\frac{A_2}{\sqrt{1+ A^2_2+B^2_2}}\nu_e
+\frac{B_2}{\sqrt{1+A^2_2+B^2_2}} \nu_\mu +
\frac{1}{\sqrt{1+A^2_2+B^2_2}}\nu_\tau,\crn \nu_2&=&
-\frac{A_1}{\sqrt{1+A^2_1+B^2_1}}\nu_e-\frac{B_1}{\sqrt{1+A^2_1+B^2_1}}
\nu_\mu- \frac{1}{\sqrt{1+A^2_1+B^2_1}}\nu_\tau,\crn
\nu_3&=&\fr{1}{\sqrt{f^2_{e\mu}+f^2_{e\tau}+f^2_{\mu\tau}}}
(f_{\mu\tau}\nu_e-f_{e\tau}\nu_\mu+f_{e\mu}\nu_\tau).\label{root}
\eea
Some useful relations are in order \cite{lvZB}
  \bea A_1A_2+B_1B_2+1&=&0,\crn A_1-\epsilon
B_1+\epsilon'&=&0,\crn A_2-\epsilon B_2+\epsilon'&=&0,\crn
(A_1-A_2)/(B_1-B_2)&=&\epsilon. \label{ABrelation}\eea One also
has \bea A_1A_2&=&\frac{(\epsilon'^2-\epsilon^2)\om'_{\mu\tau}
+\epsilon\epsilon'(\om'_{\tau\tau}-\om'_{\mu\mu})}{\epsilon\epsilon'\om'_{\mu\mu}+(1+
\epsilon^2)\om'_{\mu\tau}},\crn
B_1B_2&=&-\frac{(1+\epsilon'^2)\om'_{\mu\tau}
+\epsilon\epsilon'\om'_{\tau\tau}}{\epsilon\epsilon'\om'_{\mu\mu}+(1+\epsilon^2)\om'_{\mu\tau}}. \eea
In the standard Particle Data Group (PDG)
parametrization, the neutrino mixing
 matrix ($U_{PMNS}$) can be parametrized as \cite{PDG2012}
 \be       U_{PMNS} = \left(%
\begin{array}{ccc}
    c_{12} c_{13}     & s_{12} c_{13}                    & s_{13} e^{-i\delta}\\
    -s_{12} c_{23}-c_{12} s_{23} s_{13} e^{i \delta} & c_{12} c_{23}-s_{12} s_{23} s_{13} e^{i \delta} &s_{23} c_{13}\\
    s_{12} s_{23}-c_{12} c_{23} s_{13}e^{i \delta}&-c_{12} s_{23}-s_{12} c_{23} s_{13} e^{i \delta}  & c_{23} c_{13} \\
    \end{array}%
\right) \times P, \label{Ulepg} \ee where $P=\mathrm{diag}\left(1,
e^{i\frac{\al_{21}}{2}}, e^{i \frac{\al_{31}}{2}}\right)$, and
$c_{ij}=\cos \theta_{ij}$, $s_{ij}=\sin \theta_{ij}$ with
$\theta_{12}$, $\theta_{23}$ and $\theta_{13}$ being the solar, atmospheric
 and the reactor angles, respectively,  the
angles $\theta_{ij}=[0, \frac{\pi}{2}]$.   $\delta=[0,2\pi]$ is the
Dirac CP violation phase  and $\al_{21}, \al_{31}$ are two
Majorana CP violation phases. It is to be mentioned that in our
previous work \cite{lvZB}, the Majorana has  been  included in a
simple form.

By comparing Eqs. (\ref{UnuI}) and (\ref{Ulepg}), all the
parameters in the lepton mixing matrix in (\ref{UnuI}) can be
parameterized in terms of three Euler's angles $\theta_{ij}$ as
follows: \bea
\ep&=&\frac{c_{13}\left(\frac{B_2}{\sqrt{1+A^2_2+B^2_2}}+s_{12}c_{23}\right)}{c_{12}s^2_{13}},\,\,
\ep'=-\frac{c_{23}}{s_{23}}.\ep,\\
 e^{i\frac{\al_{21}}{2}}&=&-\frac{1}{s_{12}c_{13}}\frac{A_1}{\sqrt{1+A^2_1+B^2_1}},\crn
 e^{i\frac{\al_{31}}{2}}&=&-\frac{1}{c_{13}s_{23}}\frac{\ep}{\sqrt{1+\ep^2+\ep'^2}},\crn
 e^{i\delta }&=&-\frac{s_{13}}{c_{13}s_{23} }\ep,
\eea

and two solutions with $A_2, B_2$: \bea
A_2&=&\frac{c_{12}c_{13}}{s_{12}s_{23}+ c_{12}s_{13}c_{23}}\equiv
A^+_2,\crn
B_2&=&\frac{c_{12}s_{12}c_{13}+s_{23}c_{23}(c^2_{12}c^2_{13}-1)}{s^2_{12}
+c^2_{23}(c^2_{12}c^2_{13}-1)}\equiv
B^+_2,\label{A2B2plus} \eea or \bea
A_2&=&\frac{c_{12}s_{12}c_{13}}{s^2_{12}s_{23}-
c_{12}s_{12}s_{13}c_{23}}\equiv A^-_2,\crn
B_2&=&\frac{-c_{12}s_{12}c_{13}+s_{23}c_{23}(c^2_{12}c^2_{13}-1)}{s^2_{12}
+c^2_{23}(c^2_{12}c^2_{13}-1)}\equiv
B^-_2.\label{A2B2minus} \eea
Let us consider both the solution in Eqs. (\ref{A2B2plus}) and (\ref{A2B2minus}).

\subsubsection{\label{sectionplusI} The solution with $A^+_2, B^+_2$}

 It is easily shown that, in this case, the model is consistent
because the five experimental constraints on the mixing angles and
squared mass differences of neutrinos can be respectively fitted
with all parameters of the model. Indeed, with $A_2=A^+_2,
B_2=B^+_2$ given in Eq. (\ref{A2B2plus}), taking the data in
Ref.\cite{Nuexp1} given in Tab. \ref{InvertedH}, we obtain
\bea A_1&=&-0.95944, \hs A_2=1.56394, \hs B_1=-1.01484, \hs
B_2=-0.49319,\label{AB12}\\
 \ep&=&4.83735,\hs \ep'=-3.94968,\hs \frac{\ep}{\ep'}=-1.22474, \label{eepI} \\
e^{i\frac{\al_{21}}{2}}&=&1,\hs e^{i\frac{\al_{31}}{2}}=-1,\hs
e^{i\delta}=-1.\label{emu}\eea Eq. (\ref{emu}) implies
$\al_{21}=0$, $\al_{31}=2\pi$, $\delta =\pi$, i.e, there is no CP-violation.
The neutrino mixing
matrix then takes the form:
\be U_{\nu I}=\left(%
\begin{array}{ccc}
  0.81425 & 0.55857 & 0.15811 \\
  -0.25678 &0.59082 & -0.76485 \\
   0.52064& -0.58218 & -0.62450 \\
\end{array}%
\right).\label{numix}\ee

The physical neutrino masses are obtained as \bea
m_1&=&\sqrt{\Delta m^2_{13}}=4.9295\times 10^{-2} \,
\mathrm{eV},\crn m_2&=&\sqrt{m^2_1-\Delta
m^2_{21}}=4.8516\times10^{-2} \, \mathrm{eV},\,
m_3=0,\label{m1m2m3I}\eea and the effective masses $\langle
m_{ee}\rangle, m_\beta$ governing neutrinoless double beta decay
 \cite{betdecay1, betdecay2,betdecay3,betdecay4,betdecay5} as well as
  the sum of the neutrino masses are given by:
\bea && \langle m_{ee}\rangle = \mid \sum^3_{i=1} U_{ei}^2 m_i
\mid=
0.04782 \, \mathrm{e V},\label{meeI}\\
&& m_\beta = \sum_{i=1}^{3} \mid U_{e i}\mid^2 m_i^2=0.04843 \,
\mathrm{e V}, \label{mbetaI}\\
&& m_1+m_2+m_3=0.09781 \, \mathrm{e
 V}.\label{sumnuI}\eea
As before, we assume $\om'_{\mu\mu}=\om'_{\tau\tau}=\om'$
\cite{lvZB}. Substituting $\ep, \ep'$ in Eq. (\ref{eepI}) into
(\ref{l123}) and (\ref{numassI}) yields \be
\om'_{\mu\tau}=1.07194\om',\,\,
k=-\frac{0.0200278}{\om'},\label{omk3}\ee or \be
\om'_{\mu\tau}=1.07399\om',\,\,
k=-\frac{0.0197385}{\om'}.\label{omk4}\ee Eq. (\ref{emu}) shows
that in this case one the Dirac and one Majorana phase is nonzero, however, there is no CP violation phase. Our
next step is the second case.

\subsubsection{\label{sectionminusI} The solution with $A^-_2, B^-_2$}
In this case, taking the data in \cite{Nuexp1}  given in
Tab. \ref{InvertedH}, we obtain \bea && A_1=-0.80333, \hs
A_2=2.28904, \hs B_1=-0.65043, \hs
B_2=-1.2897,\label{AB12}\\
 &&  \ep=-4.83735,\hs \ep'=3.94968,\hs \frac{\ep}{\ep'}=-1.22474, \label{eepI2} \\
&&
e^{i\frac{\al_{21}}{2}}=e^{i\frac{\al_{31}}{2}}=e^{i\delta}=1.\label{emuI2}\eea
Eq. (\ref{emuI2}) implies $\al_{21}=\al_{31}=\delta =0$. Then, the
neutrino mixing matrix is
\be U_{\nu I}=\left(%
\begin{array}{ccc}
  0.81425 & 0.55857 & 0.15811 \\
  -0.45877 &0.45225 &0.76485 \\
   0.35572& -0.69532 & 0.6245 \\
\end{array}%
\right).\label{numix}\ee Three neutrino masses are given in
(\ref{sumnuI}), and the effective masses $\langle m_{ee}\rangle,
m_\beta$ governing neutrinoless double beta decay as well as the
sum of the neutrino masses  are given by (\ref{meeI}),
(\ref{mbetaI}) and (\ref{sumnuI}). The relation between
$\om'_{\mu\tau}, k$ and $\om'$ are given in Eqs.(\ref{omk3}) and
(\ref{omk4}). Note that in this case, all  Dirac and Majorana
violation phases are vanished.

\section{Conclusion}
 \label{conc}

In this paper we have derived the exact eigenvalues and
eigenstates of the neutrino mass matrix in the Zee-Babu model in
which the most recent data on neutrino masses and mixing with
large $\theta_{13}$ are updated. For the
inverted spectrum, one phase ($\al_{3 1}$) takes the value $2 \pi$ and the Dirac phase ($\de$) is predicted to either $0$ or $\pi$, i. e, there is no CP violation in the Zee-Babu model at the two loop level. Taking into account of the effective mass governing neutrinoless
double beta decay and the sum of neutrino, we have showed that this model fits well with  the recent experimental data in inverted spectrum.  Therefore we conclude that the Zee-Babu model is fascinating
 one for neutrino physics.

\section*{Acknowledgments}
This research has received funding from the Vietnam National Foundation for Science and
Technology Development (NAFOSTED) under grant number 103.01-2014.51. 

\end{document}